\documentclass[a4paper,fleqn,usenatbib]{mnras}
\usepackage{newtxtext,newtxmath}

\usepackage[T1]{fontenc}
\usepackage{ae,aecompl}

\usepackage{graphicx}	

\usepackage{amsmath}	
\usepackage{amssymb}	

\title[Diversity of solar minima]{The seismic diversity of four successive solar cycle minima as observed by the Birmingham Solar-Oscillations Network (BiSON)}

\author[S. Basu et~al.]{Sarbani Basu${^1}$\thanks{E-mail: sarbani.basu@yale.edu}, William J. Chaplin$^{2}$\thanks{E-mail: w.j.chaplin@bham.ac.uk}, Rachel Howe$^2$, Yvonne Elsworth$^2$,
\newauthor Steven J. Hale$^2$, Eleanor Murray$^2$\\
$^{1}$Department of Astronomy, Yale University, PO Box 208101, New Haven, CT, 065208101, USA\\ $^2$School of Physics and Astronomy, University of Birmingham, Birmingham, B15 2TT, United Kingdom}

\date{Accepted XXX. Received YYY; in original form ZZZ}

\pubyear{2025}

\begin{document}
\label{firstpage}
\pagerange{\pageref{firstpage}--\pageref{lastpage}}
\maketitle

\begin{abstract}
We have used data collected by the Birmingham Solar-Oscillations Network (BiSON) to perform a helioseismic diagnosis of changes to the Sun's internal structure between four successive solar cycle minima, beginning with the minimum at the end of cycle 21 and ending with the recent minimum at the beginning of cycle 25. The unique duration of the BiSON database makes such a study possible. We used the low-degree BiSON p-mode frequencies to constrain structural changes between minima in the layers above $\approx 0.9\,R_{\odot}$. We accomplished this by examining variations in the He\textsc{II} ionisation zone signature; and by inverting the frequency differences to infer changes in the sound speed. Additionally, we employed frequency differences between various solar models that had subtle modifications to their internal structures to facilitate analysis of the observations. We find evidence for small, but marginally significant, changes in structure between different minima. The He\textsc{II} signature was larger, and the sound speed in the range $\approx 0.93$ to $0.97\,R_{\odot}$ was slightly higher, during the cycle 23/24 minimum, than during the other minima. The cycle 23/24 minimum was the deepest, as measured by proxies of global solar activity. These findings are consistent with magnetic flux levels having been lower in this minimum than the others, resulting in a higher gas pressure, higher temperatures, and higher sound speed. Our results demonstrate the potential of using asteroseismic data to perform similar analyses on other solar-type stars.

\end{abstract}




\section{Introduction}
\label{sec:intro}

Solar minima mark the transition from one 11-year cycle of the Sun's magnetic activity to the next. The magnetic field is predominantly dipolar in nature during these periods, when the Sun is at its most quiescent and surface manifestations of magnetic activity are all but absent at active latitudes. The magnetic field configuration during solar minima is important for dynamo models, and it is an input to so-called precursor models that seek to predict levels of solar activity in the following cycle (e.g., see \citealt{Pesnell2020} and \citealt{Upton2023}). 

The minimum between solar cycles 23 and 24 was unusually quiet: the 10.7 cm flux was the lowest ever measured, and there were more sunspot-free days than those recorded in previous cycles   {\citep[see e.g.,][and also archival sunspot data  {e.g., \citealt{SILSO_Sunspot_Number}\footnote{https://www.sidc.be/SILSO/datafiles}}]{sheeley2010, hathawayLRSP}}. This minimum was also of much longer duration than any other minimum in recent history, and showed other unusual features, e.g., the polar fields were much lower than those in the minima between cycles 22 and 23, and between cycles 21 and 22. The structure of the solar corona was also different from that expected during a ``normal'' minimum \citep{Judge2010, Basu2013}. {Frequencies of solar p modes of intermediate angular degree $l$ were also unusual in that they showed more variation than was seen in the activity proxies} \citep{2010ApJ...711L..84T, 2013SoPh..282....1T}. Like the minimum between cycles 23 and 24, the one between cycles 24 and 25 has also been unusually extended and quiet. One must go back over 100 years to find a minimum as long and as deep (that between cycles 14 and 15) as the last two minima   (as observed in archival sunspot number data) . Meanwhile, the preceding two minima -- at the cycle 21/22 and 22/23 transitions -- were shorter in duration and shallower in extent, falling as they did during the Sun's \emph{modern maximum}   {(e.g., see \citealt{Usoskin2017})}, which marked an extended period when levels of solar activity were high. 

Differences between the cycle 23/24 minimum and previous minima were not confined to surface and atmospheric phenomena. Helioseismic data revealed that the interior dynamics of the Sun were different too (e.g., see differences between the cycle 22/23 and 23/24 minima reported by \citealt{Antia2010, BasuAntia2010, Howe2009}), but changes in structure have received less attention, in part because the changes are expected to be small \citep[see, e.g.,][]{Basu2021}. We now have a helioseismic database that, uniquely, stretches back over the four above-mentioned solar minima. These data were collected by the Birmingham Solar-Oscillations Network (BiSON; \citealt{Chaplin1996, Hale2016}), and allow us to perform targeted analyses. In this work, we use mode frequencies extracted from the BiSON data to measure differences in solar structure between the different solar minima. 

In previous work (\citealt{Basumin2010, Basumin2012, Basumin2013}), we explored using differences in the frequencies of low-degree solar p modes extracted from BiSON data, and found tentative indications of changes between minima. \citet{Broomhall2017} later used resolved-Sun data on higher-degree p modes, also finding differences in the mode frequencies between the cycle 22/23 and 23/24 minima. Now that BiSON has data for four solar minima, we have been motivated to carry out a more thorough analysis. This allows us to investigate differences between minima that fall in epochs having markedly changed characteristics, as noted above; and that are of opposite \emph{or the same} magnetic polarity.

Our other motivation for carrying out this work is that these Sun-as-a-star data allow us to test what is potentially possible using asteroseismic data on other stars, e.g., from future extended observations by the PLATO Mission.

The layout of the rest of the paper is as follows. In Section~\ref{sec:data} we introduce the data, beginning in Section~\ref{sec:select} with how we selected BiSON time series spanning each minimum period. Section~\ref{sec:freqs} then explains how p-mode frequencies were extracted from the data. We then present our analysis of differences in the frequencies in Section~\ref{sec:anal}. We search for evidence of structural changes in two ways. In one approach, we fitted a model of the signature in the frequencies that arises from the second ionisation of Helium (He\textsc{II}) in the near-surface layers, and tested for differences in the best-fitting parameters between minima. This is an obvious signal to test, given it is conspicuous in the low-degree mode frequencies and, moreover, is known to exhibit changes over the solar cycle (e.g., see \citealt{Basu2004, Verner2006, Watson2020}). In another approach, we compared the frequencies directly with a range of models whose near-surface structures had been perturbed in different ways. We finish the paper by summarising our main findings in Section~\ref{sec:conc}.

\section{Data selection}
\label{sec:data}

\subsection{Selection of periods covering cycle minima}
\label{sec:select}

In what follows, we label each cycle minimum using the preceding and following cycle numbers, e.g., the minimum between cycles 21 and 22 is referred to as the 21/22 minimum or Min~21/22. This differs from our previous studies in \citet{Basumin2010}, \citet{Basumin2012} and \citet{Basumin2013}, where we labelled by the following cycle number only.

\begin{figure}
\centering
\centerline{\includegraphics[width=0.5\textwidth]{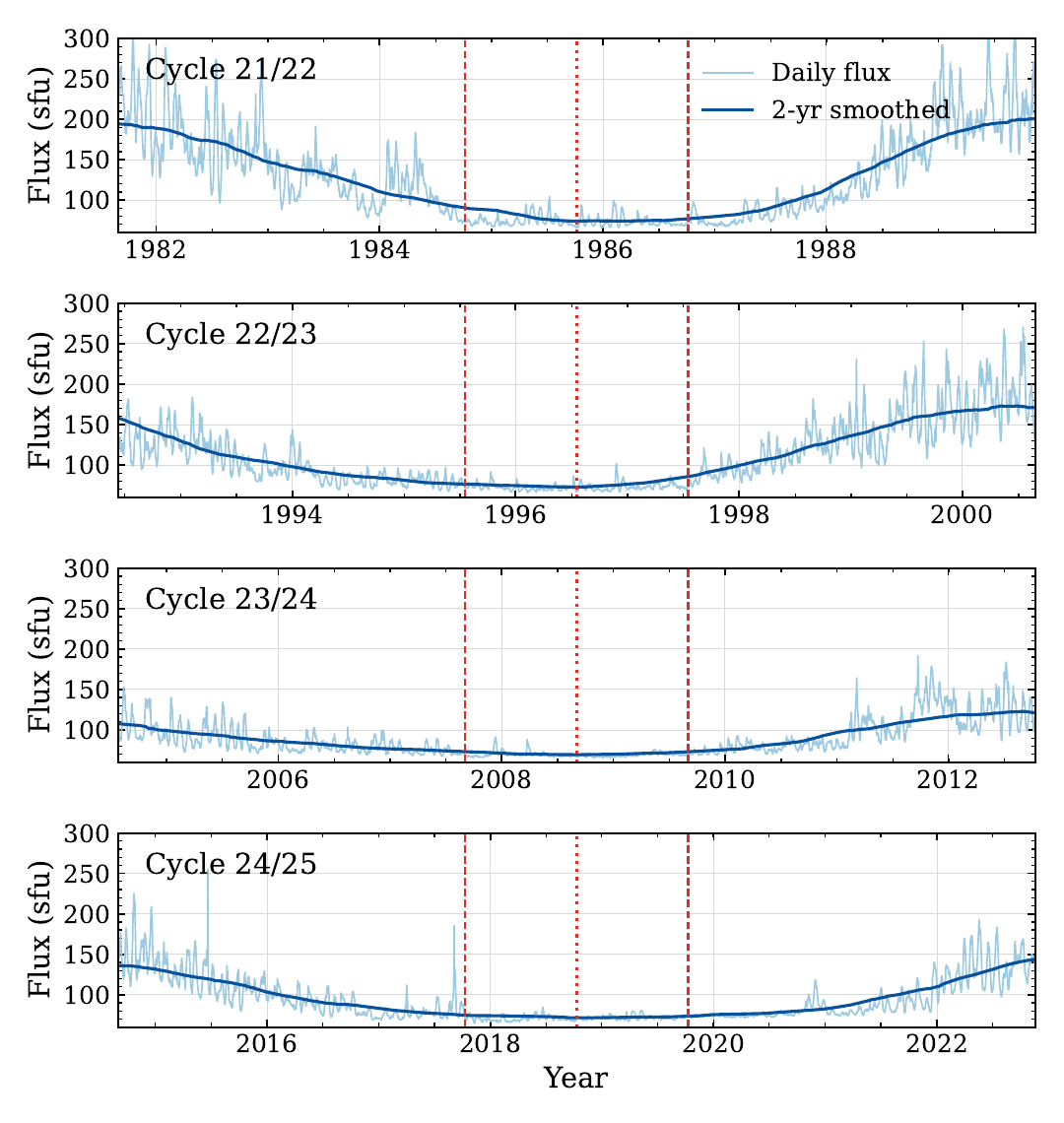}}
        \caption{Plots of the 10.7-cm radio flux spanning the four cycle minima, with daily data in grey and 2-yr boxcar averages in blue. The vertical dotted lines mark the centres of selected 2-yr periods, where the boxcar averages take their minimum values, with the start and end dates of the periods denoted by the vertical dashed lines.}
	  \label{fig:minima1}
\end{figure}

Our first step was to select the periods across each minimum for which we would extract estimates of the low-degree p-mode frequencies, from BiSON time series.  We opted for 2-yr periods as a suitable compromise between homing in on epochs when the activity is very low, which favours shorter periods and minimising uncertainties on the extracted mode frequencies which favours longer periods since the uncertainties scale to first order as one over the square root of the time series duration. We used a well-established proxy of solar activity, the 10.7-cm radio flux  {\footnote{\url{https://lasp.colorado.edu/lisird/data/penticton_radio_flux}}} (e.g., see \citealt{Tapping2013}) to fix our selection. Fig.~\ref{fig:minima1} shows the 10.7-cm radio flux across the four cycle minima, with daily data plotted in grey and 2-yr boxcar averages plotted in blue. The vertical dotted lines mark the centres of selected 2-yr periods where the boxcar averages take their minimum values. The corresponding dates are: 1985 October 6, 1996 July 18, 2008 September 2, and 2018 October 9 for the 21/22, 22/23, 23/24 and 24/25 minima, respectively. The start and end dates of each 2-yr period are marked by the vertical dashed lines. The two-year averages of the 10.7-cm radio flux  are $74.06\pm 0.24$\,sfu, $72.77\pm0.17$\,sfu, $69.33\pm   {0.12}$\,sfu and $71.24\pm 0.11$\,sfu for the 21/22, 22/23, 23/24 and 24/25 minima, respectively. The uncertainties are standard errors on each mean. The small values of the uncertainties indicate that we have selected very quiet epochs when solar activity did not change much. Note, the selection procedure adopted here differs from our previous work, where we had selected periods merely by inspection of the activity plots over time.


\subsection{BiSON data and mode frequencies}
\label{sec:freqs}

Our helioseismic diagnostics of the cycle minima -- the frequencies of the ``whole Sun'' low-degree p modes -- come from observations made by BiSON. Its database now spans five successive cycles, i.e., cycle 21 through to the current cycle 25.   {BiSON comprises six robotic telescopes at sites across the globe, which make Sun-as-a-star observations of the Sun in Doppler velocity. Data collected at each of the sites were combined using well-established techniques, as described in \citet{Davies2014}, to give multi-site timeseries for each of the 2-yr cycle minimum periods}.   {Frequencies of modes of angular degree $l=0$ to 3} were then extracted from the frequency power spectrum of each time series using the Bayesian fitting algorithm described by \cite{Howe2023},   {which fits modes a pair at a time}.   {Note the BiSON duty cycles of the 21/22, 22/23, 23/24 and 24/25 minimum periods were, respectively, 25, 77, 84, and 61 per cent. The 21/22 minimum period spanned an epoch when BiSON was not a fully deployed network, hence the much lower duty cycle. The duty cycle over the 24/25 minimum period was impacted by the global COVID-19 pandemic. It is however still possible to extract robust frequencies over the main part of the p-mode spectrum even for lower duty cycles ( {e.g., see \citealt{Elsworth1994, Basu2012, Howe2017}).}

In addition to extracting frequencies for the four 2-yr cycle minimum periods -- which constitute the principal seismic diagnostics for this study -- we also extracted frequencies for overlapping 1-yr periods spanning the entirety of the database    {(each offset by 3\,months)}. Frequency shifts were then calculated for each of the 1-yr periods, following the procedure outlined in \citet{Howe2015}. These 1-yr shifts helped us to consider the behaviour of the minima in the wider context of the full cycle trends (as we go on to discuss below).

\section{Analysis and results}
\label{sec:anal}


\begin{figure}
\centering
\includegraphics[width=0.5\textwidth]{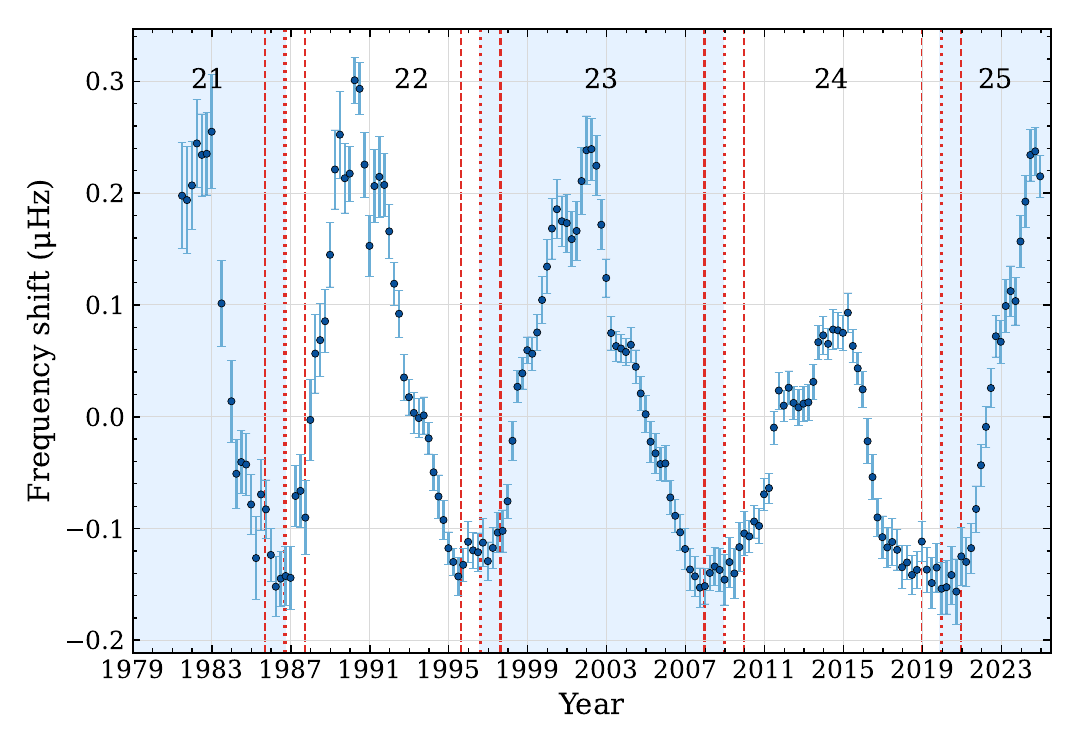}
        \caption{The average shift in the low-degree mode frequencies with respect to the temporal mean, plotted as a function of time. The coloured bands demarcate the different solar cycles, while the red vertical lines mark the minimum epoch periods flagged in Fig.~\ref{fig:minima1}.}
	  \label{fig:solcycs1}
\end{figure}

In Fig.~\ref{fig:solcycs1} we show the shift in frequencies across the solar cycles for which BiSON data exist. The shifts, determined from the overlapping 1-yr sets discussed above, were calculated with respect to the temporal mean of the frequencies for all modes in the frequency range 2500 to $3500\,\rm \mu Hz$. The figure shows clearly that the frequencies change with the solar cycle --- the frequencies are larger at solar maximum than at solar minimum. However, we also see a shorter timescale variation, which we know as the ``quasi biennial'' variation \citep{biennial}. This variation affects the frequency shifts at all epochs, and can be seen in the intervals that we have selected for this study.   {A visual inspection shows} the peak of this 2-yr signature in the cycle 23/24 minimum. There is also a hint of one in the previous cycle 22/23 minimum, and the falling phase of such a peak in the 24/25 minimum. The cycle 21/22 minimum is a curiosity --- it shows higher levels of the solar activity proxy than the other three minima, yet its frequency shifts are at least as deep as those in the cycle 23/24 and 24/25 minima. This minimum also looks very sharp in terms of the variation of the frequencies, which probably implies that the minimum of the 2-yr signature coincided with the minimum of the solar cycle. It should be noted that we have not removed the quasi-biennial signature from the frequencies that we will use for this work. Although well studied \citep[see e.g.,][etc.]{Broomhall2012biennial, Simoniello2013, Mehta2022biennial, Jain2023biennial}, the detailed nature of the signal is not fully characterised, or understood, which means we lack a sufficiently detailed description to model its robust removal from the full frequencies. 


\begin{figure}
\centering
\centerline{\includegraphics[width=3.4 true in]{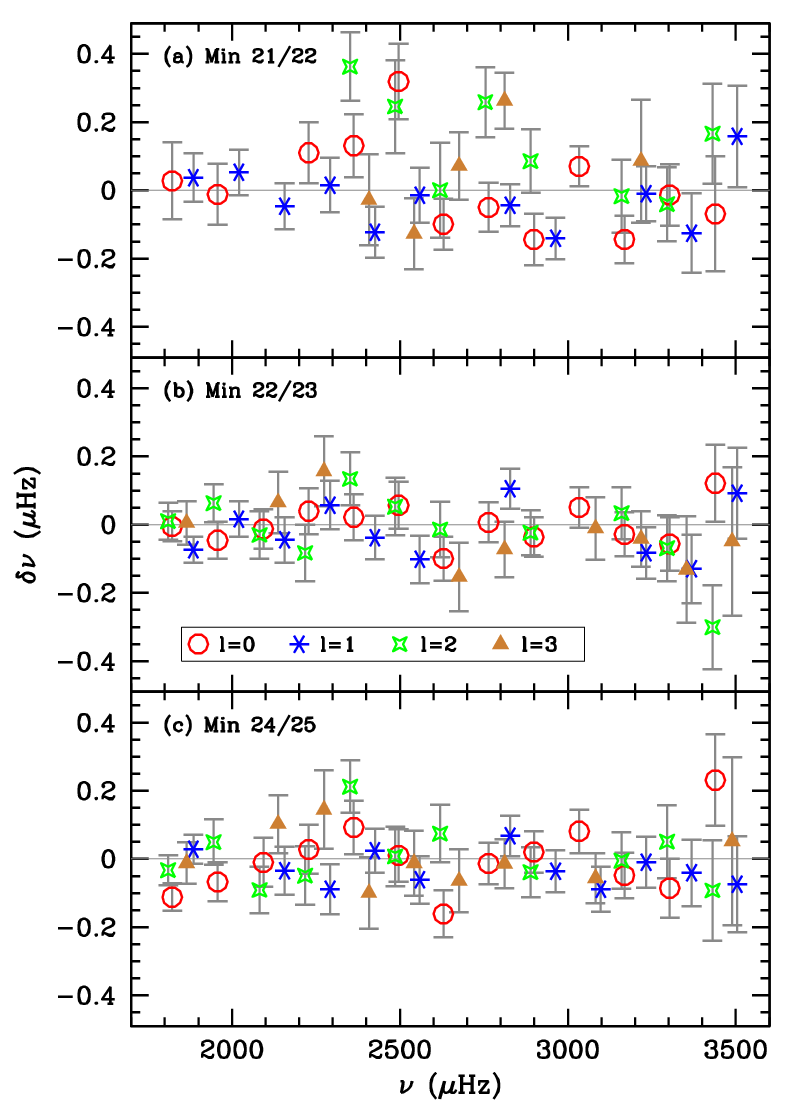}}
        \caption{The frequency differences   for each overtone $n$ and angular degree $l$ between the three minima indicated in the different panels, and those of the Cycle 23/24 minimum.   The error bars come from propagating the uncertainties on the frequencies used to construct each difference.}
	  \label{fig:freqdif}
\end{figure}


The uncertainties in the frequencies of the 23/24 minimum   {have the lowest median value}, and hence, we use those frequencies as our reference set.   {For each set, we first removed all modes that had uncertainties that were more than three times larger  than the median uncertainty of modes of the same angular degree $l$ in the above-mentioned frequency range. We also removed modes that were clear outliers in the frequency differences between the data and multiple standard solar models; these outliers were most likely the result of mode fits being biased for lower signal-to-noise realizations. This step was needed to make the subsequent analysis more robust.} In Fig.~\ref{fig:freqdif} we show the frequency differences   {for each overtone $n$ and angular degree $l$} between the other three minima and the 23/24 minimum. As can be seen, the frequency differences are very small, and a visual inspection is inadequate for evaluating whether the differences are non-random and hence significant.

To obtain a robust estimate of the significance of the differences, we performed a Wald–Wolfowitz runs test \citep{ww} on the frequency differences shown in Fig.~\ref{fig:freqdif} to check for randomness.  Since our results have uncertainties, the results from a single runs test are not adequate. For each set of frequency differences, we generated multiple sets by adding random realizations of the uncertainties to the data at each epoch.  We also created a truly random set of results at each epoch; each data point in the random sets has a Gaussian distribution with a $\sigma$ corresponding to the uncertainties of the corresponding point in the real data set. For each sequence of the realization of results, as well as the random set, we performed the runs test to derive a distribution of runs. We then compared the distribution of the runs test from our results and the random sequence. We conducted the usual two-sample KS test, as well as a Student's t~test and an F~test, to examine the null hypothesis that the distribution of runs in our results is drawn from the distribution of runs for the sample whose time variation is random. Because we are essentially dealing with small-number statistics -- in terms of the number of data points, in other words the number of runs we can get -- we relied on multiple measures of differences between the distributions of runs in the results and the distribution of runs in the random sample. The frequency differences between the 24/25 and 23/24 minima are significant at the $2.5\sigma$ level; the 22/23 and 23/24 pair at the $1.7\sigma$ level; while the 21/22 and 23/24 pair is significant at just under the $2\sigma$ ($1.93\sigma$ to be exact). Thus, we are dealing with small signals.

We proceeded with the analysis in three steps. We started with the analysis of the helium signature, for each minimum separately. Next, we sought to determine whether the observed frequency differences between the minima  could be fitted by using the frequency differences between two models. Finally, we performed what we call ``pseudo inversions'' of the observed frequency differences to determine whether the differences between two minima can be described by a difference in structure.


\begin{figure}
\centering
\centerline{\includegraphics[width=3.4 true in]{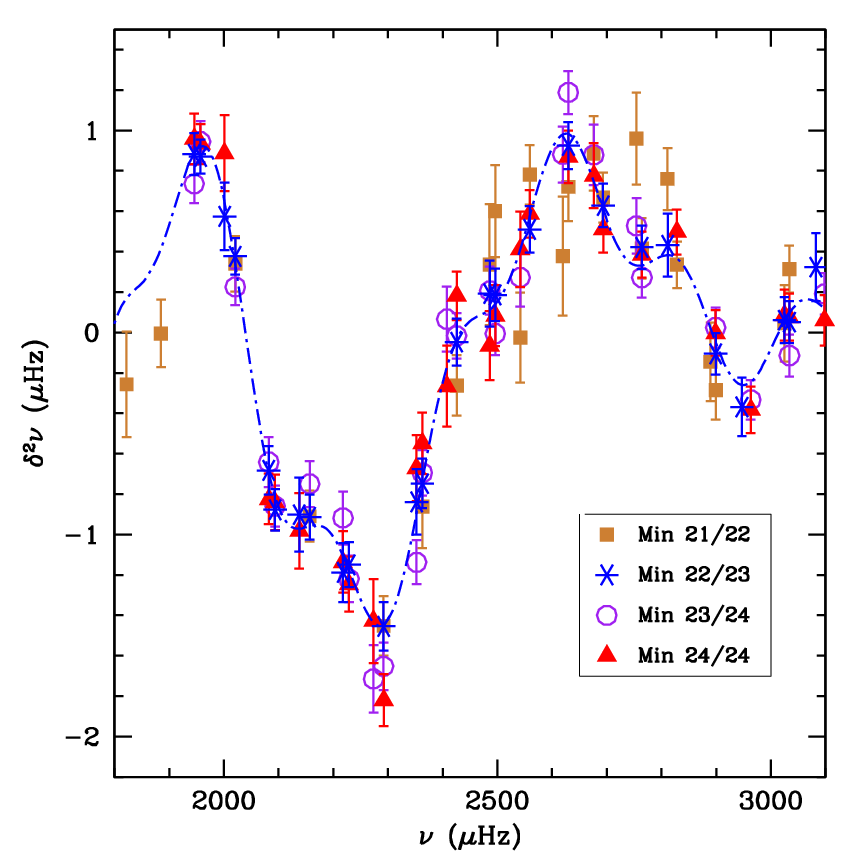}}
        \caption{The second differences of the frequencies of the different minima showing the signatures of the acoustic glitches caused by the base of the convection zone and the He\textsc{II} ionization zone. The different colours and type  of points are for the different minima as indicated in the legend. We also show the fit to the data for the 22/23 minimum as the dot-dashed line. We do not show the fits to the other sets to avoid making the figure crowded and hence, unclear.} 
	  \label{fig:2nd}
\end{figure}


\subsection{Analysis of the Helium Glitch Signature}
\label{sec:glitch}


\begin{figure}
\centering
\centerline{\includegraphics[width=3.4 true in]{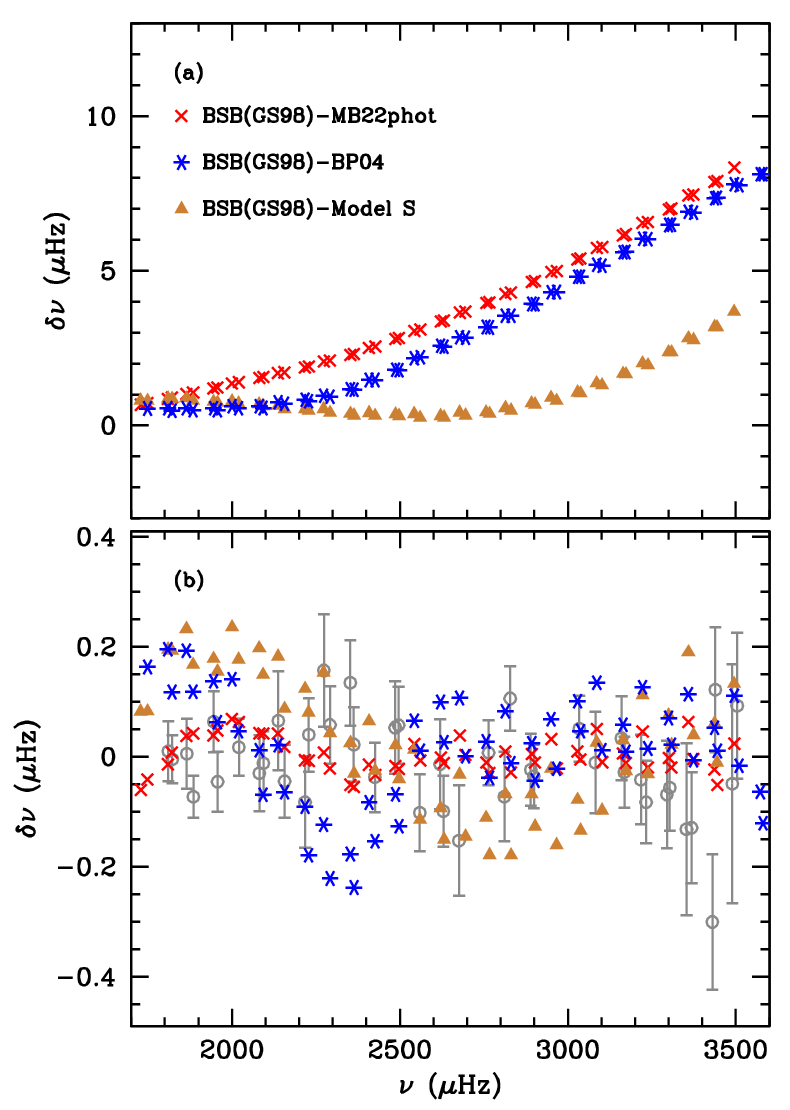}}
        \caption{Panel (a): The frequency differences of $l=0$--$3$ modes between standard solar model BSB(GS98)\citep{BSB} and standard solar models Model~S\citep{modelS}, BP04\citep{bp04}, and MB22phot \citep{mb22}. Panel (b): The same as above, but with a fit to the first   {five} Legendre polynomials   {(i.e., $P_i, i=0,4$)} subtracted out. The colours are for the same model pairs as in Panel (a). The grey points with the error-bars are the differences shown in Panel (b) of Fig.~\ref{fig:freqdif}.  }
	  \label{fig:freqdif_models}
\end{figure}


Any spherically symmetric, localized sharp feature or discontinuity in the Sun’s internal structure leaves a definite signature on the solar p-mode frequencies. \citet{Gough1990} showed that abrupt changes of this type contribute a characteristic oscillatory component to the frequencies of the modes that penetrate below the perturbation. The amplitude of the oscillatory component increases with increasing magnitude of the discontinuity, and the ``wavelength'', i.e., the corresponding period in frequency, of the component is essentially the acoustic depth of the sharp feature. These perturbations are often called acoustic glitches. For the Sun, the  He\textsc{II} ionization zone is one such feature, the other being the base of the convection zone. 
  {Since low-$l$ modes travel all the way from the surface to deep interior, their frequencies are sensitive to both acoustic glitches. Low-$l$ modes have already been used successfully to characterize the HeII glitch both in the Sun \citep{Verner2006} and in other stars \citep{mazumdar}.}

As noted earlier, several analyses \citep{Basu2004, Verner2006, Watson2020} have demonstrated that the amplitude of the He\textsc{II} acoustic glitch is sensitive to solar activity, and in particular, the amplitude decreases with increasing activity. Large magnetic fields distort the radial profile of the He\textsc{II} ionization region \citep{rings2004} and thus the change in the amplitude of the glitch can be explained as a change in the temperature profile of that region of the Sun \citep{Watson2020}, since   {there are no mechanisms that could change the helium abundance on the timescales of interest here. Gravitational settling is the only mechanism that could substantially change the helium abundance in the convection zone, but it takes hundreds of millions of years to show up as an abundance difference in stars like the Sun. }  

The oscillatory signature can be amplified by taking the second differences of the frequencies, i.e., 
\begin{equation}
\delta^2\nu_{n,l}=\nu_{n+1,l}-2\nu_{n,l}+\nu_{n-1,l},
\label{eq:2nd}
\end{equation}
where $l$ is the degree of the mode and $n$ the radial order. The second differences of the different sets of observed frequencies are shown in  Fig.~\ref{fig:2nd}. The dominant, higher-amplitude feature is the signature of the He\textsc{II} ionization zone, while the smaller-amplitude feature is the signature of the base of the convection zone.  We determined the amplitude of the glitch signature by fitting the second differences to the model   {shown in Eq.~4} of \citet{model2004}.    {We define the amplitude as the average amplitude in the frequency range of 2500--3500$\,\mu$Hz. We fitted all modes of angular degree $l$ of 0--3 that survived the weeding process described earlier.} The fit to the data for the 22/23 minimum is shown in Fig.~\ref{fig:2nd}. The uncertainties in the fitted parameters were determined through a Monte Carlo simulation by generating 2000 realisations of the data.

We found an amplitude of $0.810\pm0.044$\,$\mu$Hz for the 21/22 minimum, $0.850\pm 0.029$\,$\mu$Hz for 22/23, $1.033\pm 0.027$\,$\mu$Hz for 23/24, and $0.872\pm 0.036$\,$\mu$Hz for the 24/25 minimum.   {The 23/24 minimum amplitude is more than 3.5$\sigma$ higher than the other minimum amplitudes (based on the combined uncertainties). The other amplitudes are all consistent at the $1\sigma$ level.}   It is worth noting that the 23/24 minimum had the lowest activity, hence its larger amplitude compared with the other minima is consistent with previous studies. The lower amplitude of the 21/22 minimum signature also agrees with its higher activity. It is interesting to note that although the frequency shifts at the 21/22 minimum were as deep as the other ones (Fig.~\ref{fig:solcycs1}), this did not translate to a larger amplitude. Since one could attribute the deep minimum in the frequency shifts as being due to the minimum of the quasi-biennial signature coinciding with the solar-cycle minimum, our result could imply that the origin of the quasi-biennial signature lies closer to the surface than the helium ionization zone.

\subsection{Comparison with models}
\label{subsec:models}

One may argue that if the frequency differences between different minima are caused by changes in structure, we should be able to reproduce them by looking at frequency differences between solar models. In Fig.~\ref{fig:freqdif_models}(a) we show the frequency differences between four standard solar models,   { Model~S \citep{modelS}, BP04 \citep{bp04},  BSB(GS98)  \citep{BSB} and MB22phot \citep{mb22}. These models were chosen because they are easily available in the literature and encompass a wide range of physics inputs. For instance, Model~S was constructed with an older version of the OPAL equation of state  \citep{OPAL96} and OPAL opacities \citep{OPACold} and $Z/X=0.0245$ \citep{AN1993}. Model BP04 is a newer version of model BP04 of \citet{BP04old}, constructed with the N$^{14}$(p,$\gamma$)O$^{15}$ reaction rate of \citet{formicola}. The model also uses updated OPAL opacities \citep{OPALopacnew} and an updated OPAL equation of state, and assumes $Z/X=0.023$ \citep{GS98}. Model BSB(GS98) was constructed with $Z/X=0.023$ \citep{GS98}, OP opacities \citet{OP} and an updated OPAL equation of state \citep{opal2002}. Model MB22phot used the heavy element abundances from Table~5 of \citet{mb22}, and was constructed with updated OP opacities \citep{OPnew}, and the FreeEOS \citep{free} equation of state.} 
  {As can be seen from the figure}, the differences do not look anything like the observed differences. However, if we fit a series of low-order polynomials to the raw model differences, the resulting residuals look a lot more like the observed differences, as can be seen in Fig.~\ref{fig:freqdif_models}(b). Specifically, here we removed the first   {five} Legendre polynomials,   {$P_0$--$P_4$,} of such an expansion, as described in Appendix~A, which results in residuals that look like the observed sets. The removal of the 4th-order polynomial removes the signature of structural differences in the very near-surface layers of the model. We have explained why that is so in Appendix~A.


\begin{figure}
\centering
\centerline{\includegraphics[width=3.0 true in]{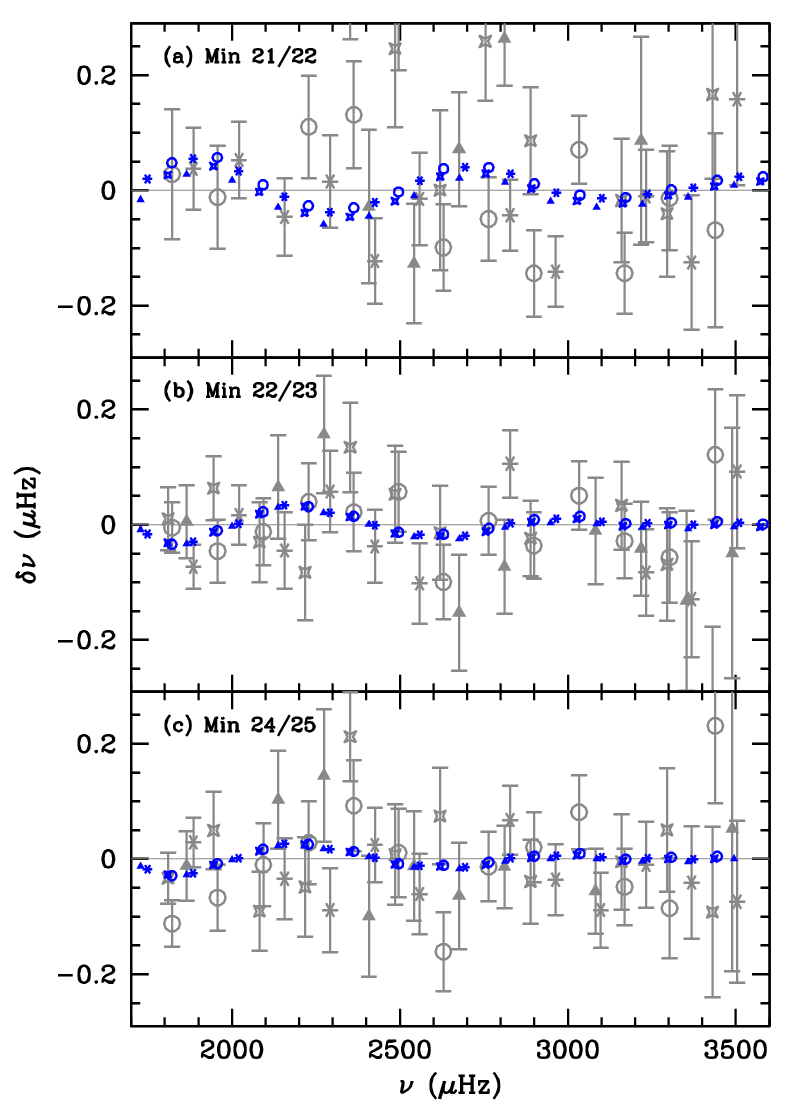}}
        \caption{The frequency differences between different minima and those of the Cycle 23/24 minimum are plotted in the different panels as grey points.   As in Fig.~\ref{fig:freqdif}, the circles, asterisks, crosses, and triangles denote differences of $l=0$, 1, 2 and 3 modes respectively.  The blue points are differences in frequencies between a modified model and a  standard solar model ; like in Fig.~\ref{fig:freqdif_models}(b), a fit to the first  five Legendre polynomials, {$P_0$--$P_4$,} have been subtracted out from the model frequency differences.  The model pair whose differences are shown was chosen such that the frequency differences between the blue and grey points had a $\chi^2$ per degree of freedom of around one.  { In panel (a), the non-standard model was constructed with OPAL equation of state corresponding to a metallicity $Z_{\rm EOS}=0.010$ in contrast to the standard model constructed with OPAL  for a metallicity of $Z_{\rm EOS}=0.018$; the panel (b) non-standard model is the OPAL equation of state for $Z_{\rm EOS}=0.025$ was used, and in panel (c), the non-standard model was constructed with the equation of state for $Z_{\rm EOS}=0.020$. } It should be noted that the fit is by no means unique, and is merely representative.}
	  \label{fig:freqdif_comp}
\end{figure}

\begin{figure}
\centering
\centerline{\includegraphics[width= 2.75 true in]{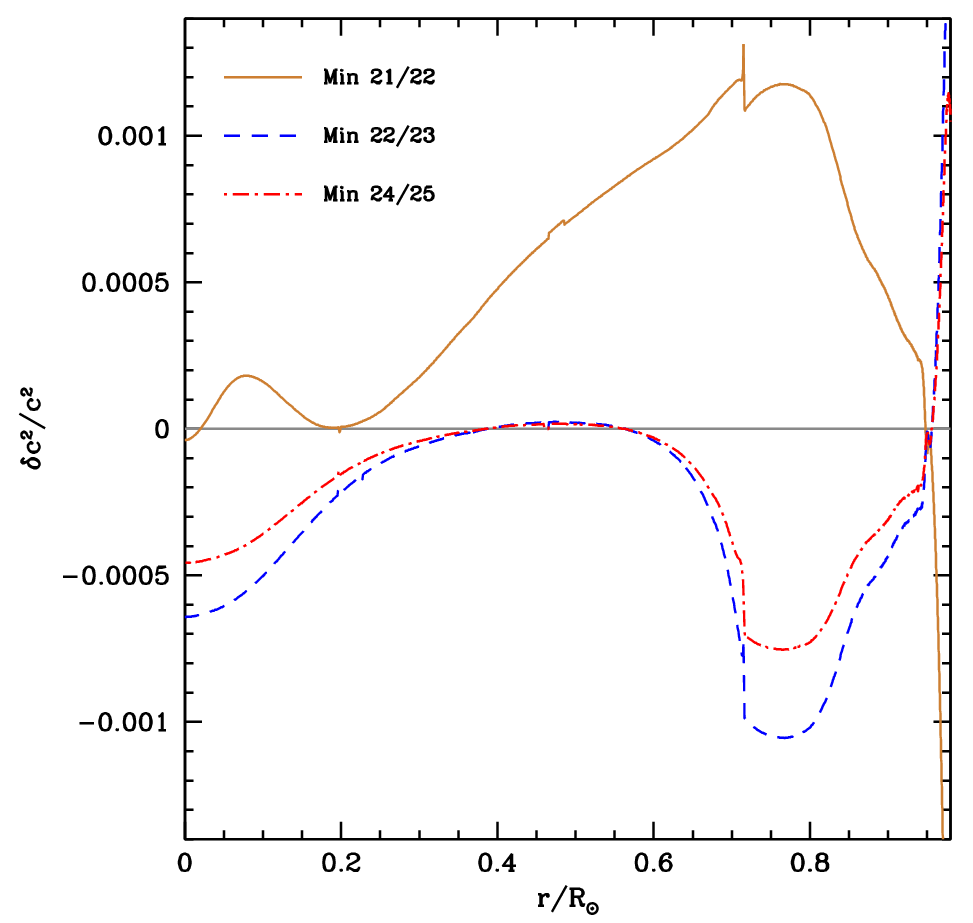}}
        \caption{The relative sound speed difference between the model pairs whose frequency differences are shown in Fig.~\ref{fig:freqdif_comp}. The different colours represent model pairs that fit the different minima. i { The orange line shows differences between the non-standard model constructed with the OPAL equation of state corresponding to a metallicity $Z_{\rm EOS}=0.010$ and a standard model ($Z_{\rm EOS}=0.018$. The non-standard model represented by the blue dashed line has $Z_{\rm EOS}=0.025$, and the red dot-dashed line is for a non-standard model with $Z_{\rm EOS}=0.020$}. The steep difference close to the surface is due to differences in the HeI+HI ionization zones arising from differences in the adopted equations of state.}
	  \label{fig:csq_models}
\end{figure}


Since it is possible to get frequency differences between models that are similar to those between the solar minima, we have tried to fit the observed differences with frequency differences of models in order to estimate the magnitude of structural changes that are required to explain the observations. The changes in the glitch signatures imply that there are changes in the convection zone, which is extremely difficult to simulate. Changes in the near-surface regions of the convection zone can be simulated by changing the helium and heavy-element abundance; however, there is no reason, or mechanism, by which solar abundances could change   {over the time scale covered by the BiSON data}. The change in the He\textsc{II} ionization profile caused by   {magnetic} activity points to changes in the equation of state, and indeed the addition of magnetic pressure would give rise to such changes. For this work, we simulated these changes by using an equation of state   {that assumes a heavy-element abundance different from the reference standard model.} These modified models were constructed in exactly the same manner as the reference standard model using the Yale Stellar Evolution Code \citep[YREC;][]{yrec}.   {The standard model was constructed with $Z/X=0.023$, OPAL opacities \citep{OPALopacnew}, and the OPAL equation of state \citep{opal2002}. Gravitational settling of elements was calculated using the coefficients of \citet{thoul}.}   { We calculated the frequencies of the models using the pulsation code of \citet{Antia_puls}. We then } determined the frequency differences with respect to a baseline standard model, fitted and removed the first   {five} Legendre polynomials   {i.e., $P_0$--$P_4$}, and then compared the residuals with the observations. The results can be seen in Fig.~\ref{fig:freqdif_comp}. The relative differences in the squared sound-speed differences between the models are shown in Fig.~\ref{fig:csq_models}. 

Unfortunately, the solutions we find are not unique. Many similar changes can mimic the frequency differences --- for example, one could use an equation of state of a different metallicity, as we have done, or make models including magnetic fields such as those of \citet{Lin2009}, or try completely different equations of state. Besides, the work of \citet{Basu2021} suggests that the actual differences in structure are much smaller than the differences plotted in Fig.~\ref{fig:csq_models}.  Thus, the results in Fig.~\ref{fig:freqdif_comp} and Fig.~\ref{fig:csq_models} are merely illustrative, and show that the frequency differences can be explained by structural differences. To get better estimates of the differences, we turn to the inversions of the frequency differences.

\subsection{Inversions of the frequency differences}
\label{sec:inv}


\begin{figure}
\centering
\centerline{\includegraphics[width=3.25 true in]{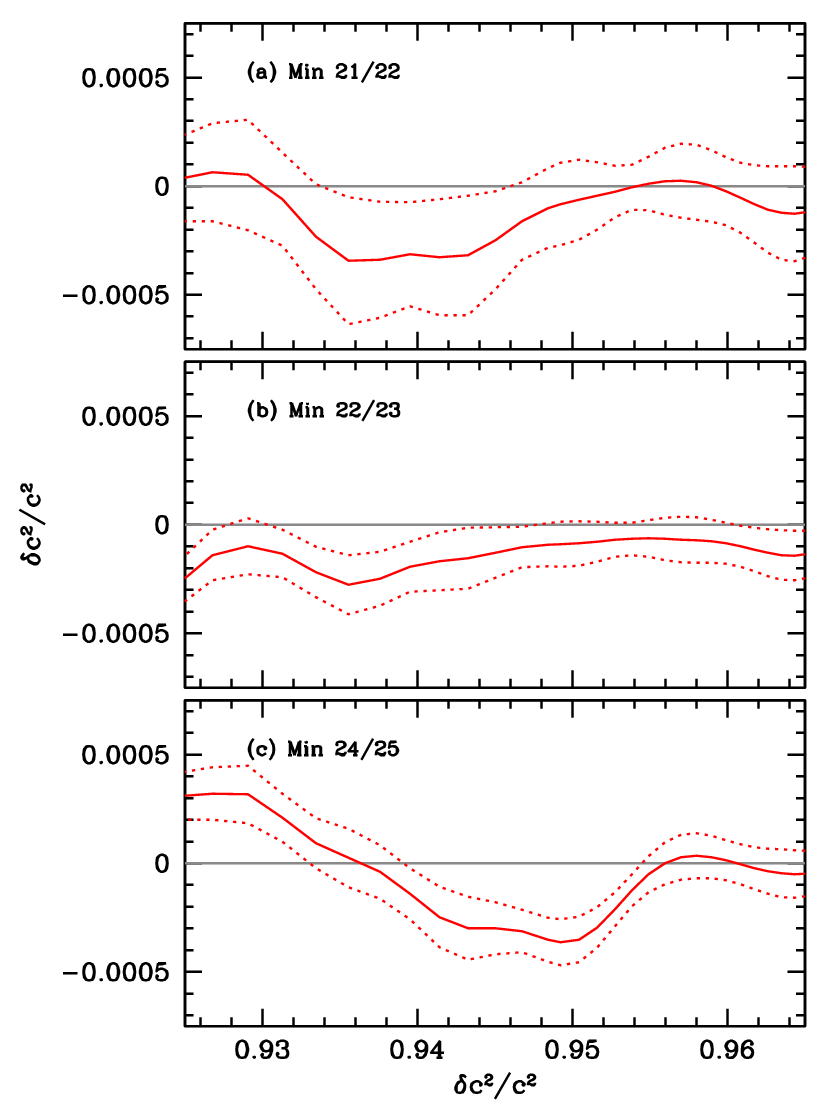}}
        \caption{The relative difference in the squared sound-speed between the different minima and the 23/24 minimum. The results were obtained from the frequency differences shown in Fig.~\ref{fig:freqdif}.   {The uncertainties are the standard deviation obtained by inverting the 2000 realizations of the data.}}
	  \label{fig:inv}
\end{figure}


Inversions of frequency differences are an established method to determine the difference in structure between the Sun and a solar model \citep[see, e.g.,][]{lrsp}. Inversions using the differences in observed frequencies between two different epochs \citep{Basu2021} or two local regions of the Sun \citep{rings2004} have also been made successfully. We, however, face the challenge of a very limited dataset -- i.e., one that only includes low-degree modes -- and have to use a different method of carrying out the inversions. We have described the technique in Appendix~B. Our inversion results are shown in Fig.~\ref{fig:inv}.

The inversions reveal small differences in the structure of the Sun between the 21/22, 22/23, 24/25 minima and the 23/24 minimum. The differences are predominantly negative, which is what we would expect given that the 23/24 minimum was quieter than the other minima.   {The sound-speed differences do not behave monotonically with the differences in the 10.7cm flux; we speculate that this is a result of the quasi-biennial variation of the frequencies being at different phases in the different minima, even though this signal does not affect the HeII signature at the shallower radius of $0.98$ R$_\odot$.}

The sound-speed differences   {that we find} are small, but larger than those that \citet{Basu2021} found at the base of the convection zone. Of course, the base of the convection zone has a much higher gas pressure than the more superficial layers covered by our inversions here, and a very large magnetic field is required to change the structure there (the change depends on the ratio of the magnetic pressure to gas pressure).   {Closer to the surface, where gas pressure is small, one would expect slightly larger differences. Indeed, we find that in the radius range $0.92$-$0.97$ R$_\odot$, the differences are larger than at the convection-zone base.}
 However, it should again be borne in mind that the data we use here are much more limited. For reference, using data for modes of degree $l=0$ to $l=150$ obtained by the Global Oscillation Network Group GONG project, \citet{Basu2021} found a 
relative squared sound-speed   {difference} of $(1.05 \pm 0.29)\times10^{-5}$   {for the 22/23 minimum with respect to the  23/24 minimum} in the radius range $0.75\pm 0.03R_\odot$; the difference was smaller, i.e., $(0.25 \pm 0.30)\times10^{-5}$, for the 24/25 minimum   {when compared with the 23/24 minimum}. GONG started observing in May 1995, and hence there are no data on the 21/22 minimum. Note that \citet{Basu2021} used the GONG pipeline data that are analysed in lengths of 108 days; the numbers quoted above are calculated as the mean of the non-overlapping 108-day sets that cover the time-intervals in question. Unfortunately, there are no space-based data that cover all the minima in question; the Michelson Doppler Imager instrument \citep[MDI, ][]{mdi} onboard the Solar and Heliospheric Observatory (SoHO) only covered the 23/24 minimum, while the Helioseismic and Magnetic Imager \citep[HMI, ][]{hmi} onboard the Solar Dynamics Observatory covered the 24/25 minimum.   {While the Global Oscillations at Low Frequencies \citep[GOLF,][]{golf} instrument onboard SoHO has observed the 23/24 and 24/25 minima, it did not observe the 21/22 minimum at all, and its coverage of the 22/23 minimum is incomplete; besides, GOLF only obsrves low-degree modes, just like BiSON.}



\section{Discussion and Conclusions}
\label{sec:conc}

We have made use of the extremely long-duration BiSON database to enable a unique helioseismic diagnosis of changes in internal structure between four successive cycle minima. The dataset includes two pairs of minima that share the same magnetic polarity. Each pair has one minimum in the \emph{modern maximum} epoch (the minima between cycles 21/22 and 22/23, respectively), and one minimum in the lower activity epoch that has since followed (the minima between cycles 23/24 and 24/25, respectively). Whilst it is, of course, known that solar maxima can be very different, here we show that even small differences in the activity levels at cycle minima can lead to noticeable differences in internal structure between those epochs.

We used differences in the whole-Sun, low-degree BiSON mode frequencies as our structural diagnostic, and sought to constrain structural changes between minima in the layers above $\approx 0.9\,R_{\odot}$ by: looking for differences in fits to the He\textsc{II} ionisation zone signature; and inverting differences in the frequencies to infer changes in the sound speed. We also used frequency differences between different solar models, which have subtle alterations to their internal structures, to help understand and support the analysis of the real observations.   {Unlike earlier works, such as those by \citet{Basumin2010, Basumin2012, Basumin2013}, we determined the epochs of minima by careful consideration of the 10.7-cm radio flux; we also used a wider range of models to compare the data with. While earlier results had hinted that there were differences in solar structure between the different minima, here we have been able to properly quantify the differences, and have extended the analysis to include the latest cycle 24/25 minimum.}

We find evidence for small, but marginally significant, changes in structure between different minima. The He\textsc{II} signature was larger, and the sound speed in the range $\approx 0.93$ to $0.97\,R_{\odot}$ was slightly higher, during the cycle 23/24 minimum than during the other minima. The cycle 23/24 minimum had the deepest minimum as measured by proxies of global solar activity, such as the 10.7-cm radio flux. Our results are therefore consistent with levels of magnetic flux being lower in this minimum than the others, resulting in a higher gas pressure, higher temperatures, and higher sound speed. Regarding the changing magnetic polarity across the four minima, the polarity itself should not be a direct factor in determining the frequency shifts since they depend on the square of the magnetic field, not the field itself (e.g., see \citealt{dogmjt1990}). However, this does not rule out structural changes that follow a 22- as opposed to an 11-yr cycle. Here, we find no significant evidence for such changes. Finally, the low-degree mode frequencies contain signatures of the so-called quasi-biennial signature, and our analyses suggest that the origin of this signature may lie closer to the surface than the helium ionization zone.

These results demonstrate the potential feasibility of using asteroseismic data to perform similar analyses on other solar-type stars. Extended future observations in a single field by the upcoming PLATO Mission (\citealt{Goupil2024, Rauer2025}) should enable such analyses on stars with shorter cycle periods than the Sun, permitting the capture of two successive minima in a single dataset. High-quality asteroseismic targets for PLATO should yield frequencies of sufficient quality. It is also worth making the more general point that our approach here provides a template to search for structural differences between any two epochs, irrespective of whether or not they coincide with stellar cycle minima.

\section*{Acknowledgements}

We would like to thank all those who are, or have been, associated with the Birmingham Solar-Oscillations Network (BiSON), in particular P. Pallé and T. Roca-Cortes in Tenerife and E. Rhodes Jr. and colleagues at Mt. Wilson. W.J.C., R.H., Y.E. S.J.H. and E.M. acknowledge the support of the United Kingdom Science and Technology Facilities Council (STFC) through grant ST/V000500/1. S.B. acknowledges NASA grant 80NSSC25K7669. This research has made use of NASA's Astrophysics Data System Bibliographic Services. The sunspot data mentioned in this work were obtained from WDC-SILSO, Royal Observatory of Belgium, Brussels.

\section*{Data Availability}

 The BiSON time series analysed here is available at \url{http://bison.ph.bham.ac.uk/opendata}.


\bibliographystyle{mnras}
\bibliography{mnmin1} 

\appendix

\section{Implications of removing the ``surface term''}
\label{sec:surface}

Solar models constructed using the standard equations of stellar structure and evolution are very good, with only small differences in structure between the Sun and the models \citep[see, e.g.,][]{fresh}. However, there is a very large frequency-dependent (but angular degree independent) difference between the frequencies of the models and the observed frequencies. This is called the ``surface term'' because it arises from near-surface errors in modelling, such as the approximations made in describing convection, etc. Frequency differences between two models can also have such frequency differences because of differing physics and boundary conditions used; this can be seen in Fig.~\ref{fig:freqdif_models}(a) and also Fig.~\ref{fig:surface}(a). This difference needs to be removed first to understand frequency differences caused by differences in structure.

\begin{figure}
\centering
\centerline{\includegraphics[width=3.1 true in]{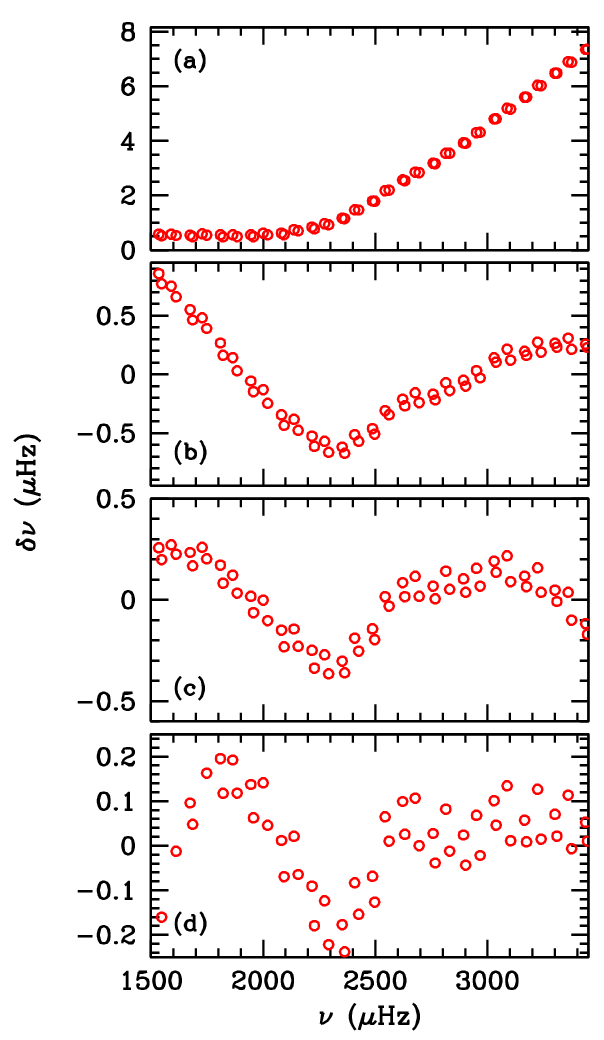}}
        \caption{Panel~(a): The frequency differences between model BSB and model BP04.   {Only differences for models of degree $l=0$, 1, 2 and 3 are shown in this and subsequent panels.} Panel~(b): The residuals after removing the \citet{bg14} two-term function from the differences shown in (a). Panel~(c): The residuals after removing the first three Legendre polynomials from the differences in (a). Panel~(d): The residuals after removing the first five Legendre polynomials. Note that the differences in Panel~(d) look like the differences shown in Fig.~\ref{fig:freqdif}. }
	  \label{fig:surface}
\end{figure}


In asteroseismic studies, the norm is to subtract out the so-called BG14 form \citep{bg14}, where the frequency differences are fitted to a form:
\begin{equation}
    \delta\nu_i=\frac{1}{I_i}\left[\frac{a_{-1}}{\nu_i}+a_3\nu_i^3\right],
\end{equation}
where the $\nu_i$ is the frequency of the $i$th mode, $\delta\nu_i$ is the frequency difference, $I_i$ is the mode inertia, and $a_{-1}$ and $a_3$ are free parameters determined from the fit. The residuals of the fit may then be used to compare differences in the structure of the models. We show in Fig.~\ref{fig:surface}(b) these residuals, after fitting this form to the differences in Fig.~\ref{fig:surface}(a). Note that the residuals do not look anything like the frequency differences between two minima shown in Fig.~\ref{fig:freqdif_comp}.

With solar data, the norm has generally been to express the surface term in terms of low-degree polynomials, such as Legendre polynomials. Fig.~\ref{fig:surface}(c) shows what happens if we remove the first four Legendre polynomials   {(i.e., $P_i$, $i=0,1,2,3$)}; the differences still do not look like the observed differences. However, removing the first five Legendre polynomials   {(i.e., $P_i$, $i=0,1,2,3,4$)} yields residuals that do look like the frequency differences between the different solar minima (Fig.~\ref{fig:surface}d). Consequently, that is what we have done with the model pairs.

To understand the implications of fitting and removing the first four Legendre polynomials from the frequency differences in Fig.~\ref{fig:surface}(a), we turn  to the formalism of \citet{filter}. That work showed that removing the surface term from the frequency differences is tantamount to removing information of the near-surface layers. 

The frequency differences between the Sun and a solar model, or between two solar models, can be written as \citep[e.g.,][]{lrsp}:
 \begin{equation}
 \begin{aligned}
 \frac{\delta \nu_i}{\nu_i}\equiv d_i
& = \int_0^{R_\odot} K_{c^2,\rho}^i(r)\frac{ \delta c^2(r)}{c^2(r)}\; {\rm d} r +
  \int_0^{R_\odot} K_{\rho,c^2}^i(r) \frac{\delta \rho(r)}{\rho(r)}\; {\rm d} r\\
 & +\frac{F_{\rm surf}(\nu_i)}{I_i},
 \label{eq:inv}
 \end{aligned}
 \end{equation}
 where $\delta \nu_i$ is the difference in the frequency $\nu_i$ of the $i$th mode between the data and the reference model. The functions  $K_{c^2, \rho}^i$ and $K_{\rho, c^2}^i$ are the ``kernels'' and are known functions that relate the changes in frequency to the changes in the squared sound speed $c^2$ and density $\rho$, respectively, and $I_i$ is the mode inertia. The term $F_{\rm surf}$ is the ``surface term''. \citet{filter} showed that removing the surface term effectively changes the kernels.

If $F_{\rm surf}$ is expressed in terms of basis functions in frequency -- in our case the Legendre polynomials, but it can be anything -- we may write it as a vector:
\begin{equation}
[{\mathbf F}]_i=\frac{F_{\rm surf}(\nu_i)}{I_i}=\sum_j\frac{\Phi_{ij}}{I_i},\quad j=1,\ldots\Lambda.
\label{eq:filt}
\end{equation}
A singular value decomposition of $\mathbf F$ can be be written as ${\mathbf F}={\mathbf U}{\mathbf \Sigma}{\mathbf B}^T$. ${\mathbf \Sigma}$ is the diagonal matrix that contains the singular values $\sigma_i$ of ${\mathbf F}$. \citet{filter} showed that equivalent filter is
\begin{equation}
{\mathbf G}={\mathbf I}-{\mathbf U}{\mathbf U}^T,
\label{eq:gmat}
\end{equation}
where ${\mathbf I}$ is the identity matrix. The matrix ${\mathbf G}$ has $\Lambda$ singular values that are equal to zero. Removing the surface term is equivalent to applying the filter ${\mathbf G}$ on ${\mathbf d}$ the vector
comprising the observations, i.e., the LHS of Eq.~\ref{eq:inv}. The components that are removed from the kernels are terms in ${\mathbf GK}$, that correspond to the zero singular values, where ${\mathbf K}$ is the matrix with the kernels. 
Since
\begin{equation}
({\mathbf G}{\mathbf d})_i=\sum_k{\mathbf U}_{ik}\sigma_k\left(\sum_j{\mathbf V}^T_{kj}{\mathbf d}_j\right),
\label{eq:sum}
\end{equation}
and if $\sigma_k=0$, the corresponding components do not contribute to the sum, they are effectively filtered out.


\begin{figure}
\centerline{\hbox{\includegraphics[width = 3.1 true in]{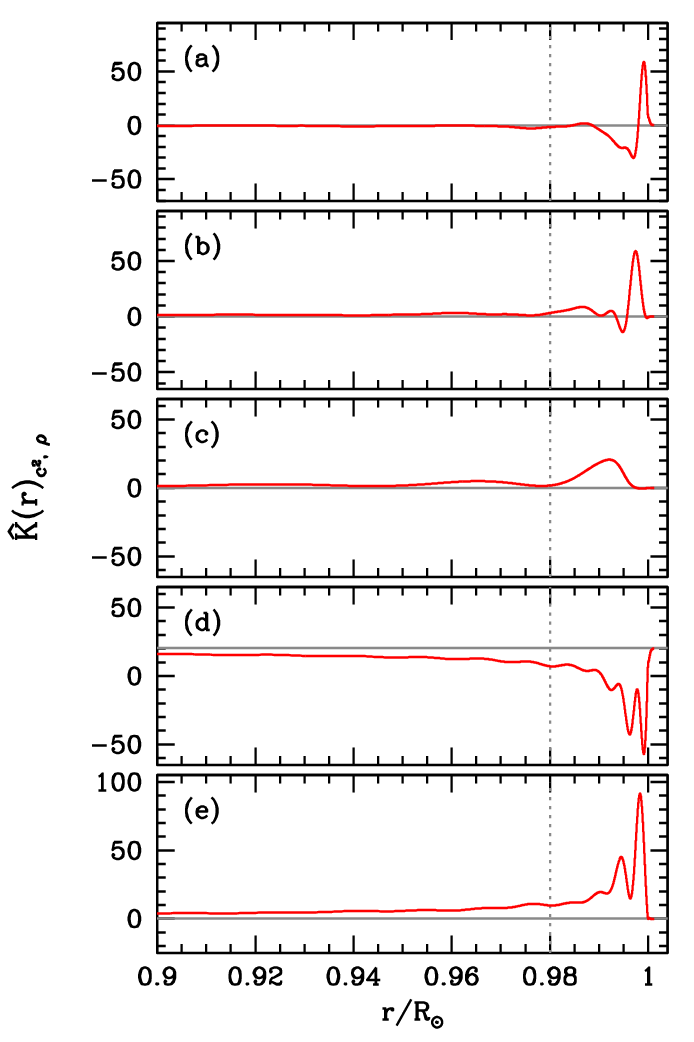}}}
\caption{The components of sound-speed kernels that are removed when the   {$P_0$--$P_4$} Legendre polynomials
were removed from the frequency differences in Fig.~\ref{fig:surface}(a).   {The five terms that were removed result in five singular values equal to 0 in the expression shown in  Eq.~\ref{eq:gmat}, which results in 5 components removed from the kernels. Each panel corresponds to one of the components removed. } Since the polynomials were removed simultaneously, one cannot make a one-to-one correspondence between the polynomial order and the component removed. The vertical line marks the approximate position of the He{\textsc{II}} ionisation zone.
}\label{fig:khat}
\end{figure}


The components of the sound-speed kernels that we have filtered out by our fit to the differences in Fig.~\ref{fig:surface}(a) are shown in Fig.~\ref{fig:khat}. As can be seen, the discarded components are non-zero and large only close to the surface and above the second helium ionisation zone, i.e., approximately $0.98$R$_\odot$. Since we have had to remove these components from the frequency differences of models to approximate the observed frequency differences between the different minima, the implication is that differences in solar structure between those epochs were not confined to the very near-surface layers.

\section{The ``inversion'' technique}
\label{sec:inv}

Usually Eq.~\ref{eq:inv} is used to determine the differences in structure between the Sun and solar models, as a means of determining both the structure of the Sun as well as how good the model is. However, a reliable inversion requires many more frequencies than we have access to here --- it is quite normal to use frequencies between 1500 and 3500 $\mu$ Hz for modes with spherical harmonic degrees from 0 to 150 or 200, see e.g., \citealt{fresh}).


\begin{figure*}
\centerline{\hbox{\includegraphics[width= 5 true in]{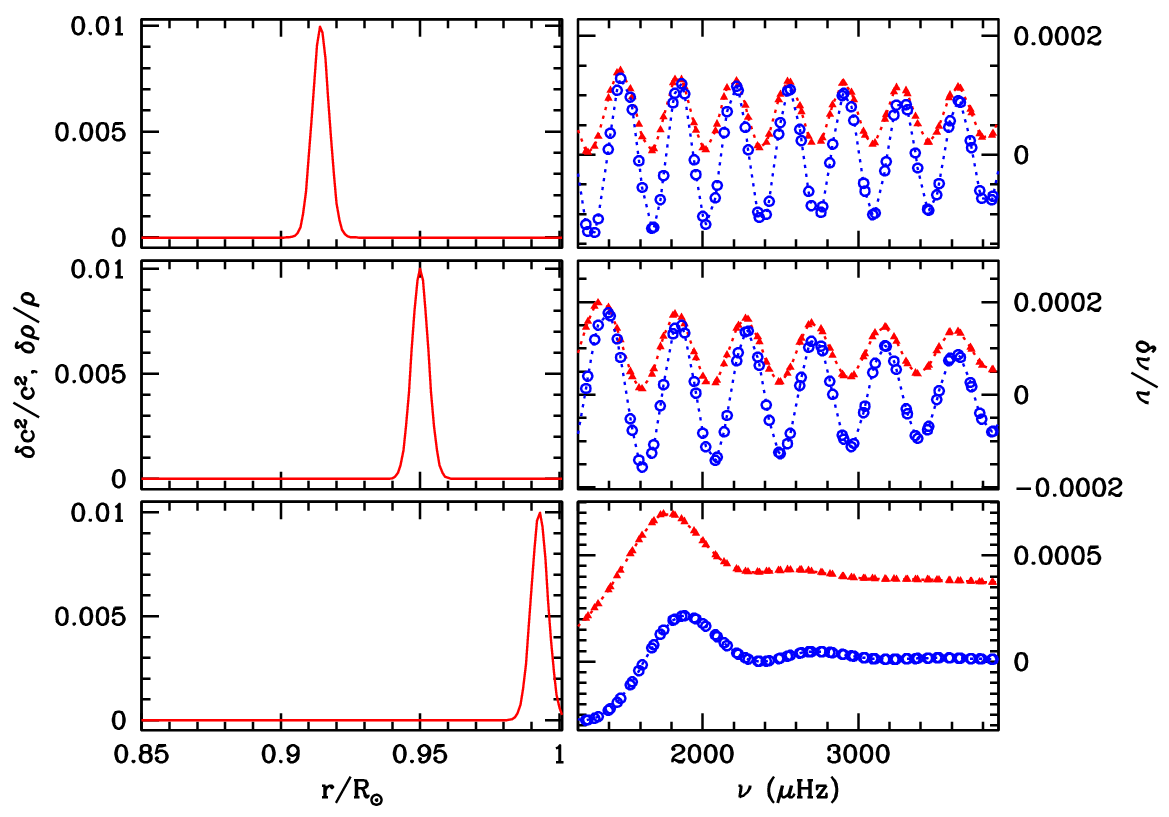}}}
\caption{Localised changes in the relative sound-speed and density differences
(left column) and the corresponding relative frequency changes (right column).
In the right column, the points in red are the changes as a result of
sound-speed changes, while the points in blue are the changes caused
by the density differences.}\label{fig:kernels}
\end{figure*}


Inverting only $l=0$,1,2 and 3 modes using Eq.~\ref{eq:inv} does not provide us reliable information except in the solar core \citep[etc.]{basu2003, lynn}.
We therefore use a hybrid forward modelling/fitting pseudo-inversion technique to determine the structure differences between the different solar minima. Since we know the kernels linking sound speed and density differences to frequency differences, we can calculate frequency differences that would be caused by localised differences in sound-speed and density. Thus, in the case of sound speed,
\begin{equation} 
{\frac{\delta\nu_i}{\nu_i}}=\int_0^{R_\odot} K_{c^2,\rho}^i(r)f(r),
\label{eq:forward}
\end{equation}
where $f(r)$ is a localised function that represents a localised relative sound-speed difference. A similar expression can be used to calculate frequency differences that correspond to density differences. We use Gaussians to represent the localised difference $f(r)$. The frequency differences obtained for a set of $f(r)$ localised at different   {radii} are then treated as basis functions to fit the frequency differences in Fig.~\ref{fig:freqdif}. A sample of some of the localised changes and the corresponding frequency differences are shown in Fig.~\ref{fig:kernels}. It should be noted that handling the sound-speed and density changes independently is not self-consistent, because a change in density will cause a change in sound-speed, and vice versa. However, since the asymptotic theory of oscillations shows that the frequency differences can be explained well in terms of sound speed alone \citep{aerts, ourbook}, we only determine sound-speed changes.   {We emphasize again that this is not a true inversion process, but a very constrained one, which allows us to use only low-degree modes to estimate sound speed differences.}


\begin{figure}
\centerline{\hbox{\includegraphics[width = 3.25 true in]{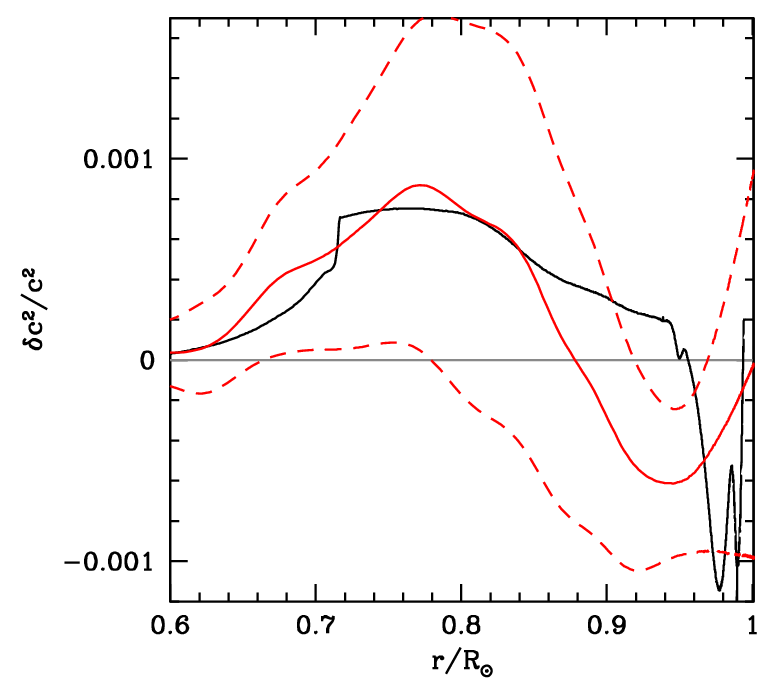}}}
\caption{Our attempt to invert the frequency differences of the model pair in Fig.~\ref{fig:freqdif_comp}(c) whose sound-speed differences are shown in Fig.~\ref{fig:csq_models}(c). The exact difference is shown as the solid black line. The inverted result is the solid red line, with 1$\sigma$ uncertainties shown as the red dashed lines. The uncertainties are based on using the uncertainties in the frequency difference between the Cycle~24/25 and Cycle~23/24 minima.}
\label{fig:wrong}
\end{figure}


\begin{figure}
\centerline{\hbox{\includegraphics[width = 3.25 true in]{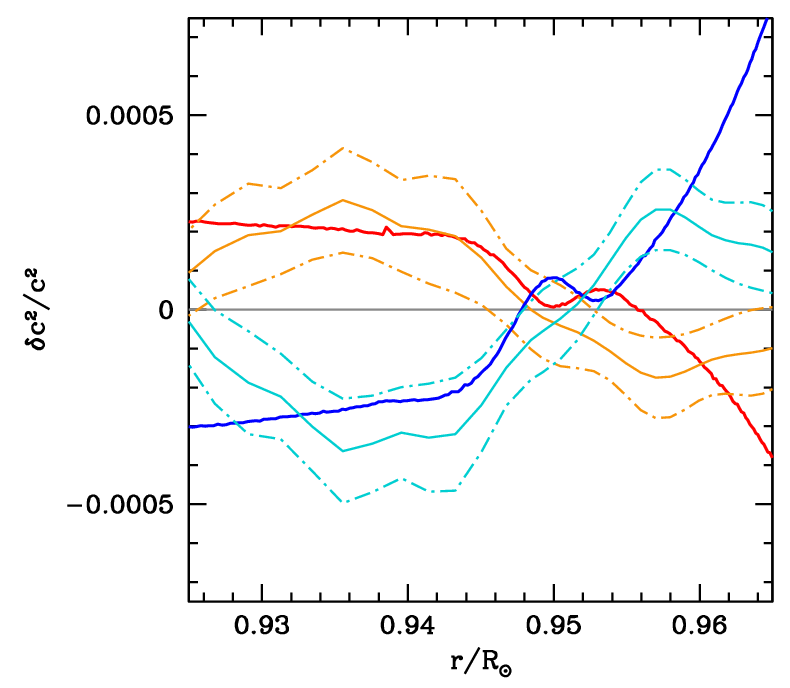}}}
\caption{Near-surface inversions of the frequency differences of the model pairs in Fig.~\ref{fig:freqdif_comp}(a) and (c)   {(whose sound-speed differences are  shown in Fig.~\ref{fig:csq_models} as the brown and red lines respectively)}. The exact differences are shown as the thick solid   {lines, red for the model pair in Fig.~\ref{fig:freqdif_comp}(a), ans blue for the model pair in Fig.~\ref{fig:freqdif_comp}(c)}. The thin solid orange line is the inversion result for the model pair shown in red, and the light-blue line is that for the model pair shown in blue. The dot-dashed lines are 1$\sigma$ uncertainties, based on using the uncertainties in the observed frequency differences between the Cycle~24/25 and Cycle~23/24 minima. Note that the inversion results cannot be trusted below $0.93R_\odot$ and above   {$0.96R_\odot$}.}
\label{fig:right}
\end{figure}


Given the paucity of our data, we cannot use too many basis functions, which means that we can only obtain low-resolution, smooth solutions. To test the robustness of our inversions, we used the frequency differences of the model pair in Fig.~\ref{fig:freqdif_comp}(c), whose sound-speed differences are shown in Fig.~\ref{fig:csq_models}. We first attempted to determine the sound-speed differences between 0.6 and $0.99R_\odot$. We added uncertainties to the frequency differences of the models to simulate observations and used 1500 realizations of the data in order to find the mean solution as well as its uncertainties. The results are shown in Fig.~\ref{fig:wrong}, and as can be seen they are not very good, with uncertainties are much too large for the results to be useful. Note that the reduction of uncertainties around $0.6R_\odot$ is an artefact of running out of the Gaussian basis functions.   {Thus, clearly, a low-resolution inversion over a large radius range is not useful.}

Since it is clear that we cannot invert for structure difference deep inside the convection zone, we concentrated instead on inverting for the structure closer to the surface, below the Helium ionization zone. As can be seen from Fig.~\ref{fig:right}, where we show the inversion results for two different pairs of models, this time we can get reliable inversion results, here between about $0.93R_\odot$ and   {$0.96R_\odot$}. Hence, we only show these near-surface inversion results in Fig.~\ref{fig:inv}.


\bsp	
\label{lastpage}
\end{document}